# THE SIGNIFICANCE OF THE BURNS TEMPERATURE IN RELAXOR PMN

S.A. Prosandeev

*Physics Department, Rostov State University, 344090 Rostov on Don, Russia*

S.B. Vakhrushev, N.M. Okuneva, and P.A. Markovin

*A. F. Ioffe Physical-Technical Institute,194021 St. Petersburg, Russia*

I.P. Raevski and I.N. Zaharchenko

*Institute of Physics, Rostov State University, 344090 Rostov on Don, Russia*

M.S. Prosandeeva

*Institute of Theoretical and Experimental Biophysics, Puschino, Russia*

We suggest considering the Burns temperature in lead containing relaxor ferroelectrics as the temperature, at which Pb becomes off-center and nearly randomly occupies positions around the center on a sphere with the radius, which increases with the temperature decrease. The significance of the Burns temperature, in this approach, is the crossover between the soft-mode and order-disorder dynamics. The lattice parameter, birefringence and neutron scattering experiments are analyzed in the light of this model.

## 1. Introduction

Relaxor ferroelectrics attracted much attention last years because of remarkable piezoelectric properties of PMN-PT, PZT-PT, LPZT-PT etc [1] and because of fundamental problems of the description of properties of locally disordered solids. The progress in the understanding of these significant properties is not that great because one of the main peculiarities of the relaxor ferroelectrics, the existence of the Burns temperature, has not been well understood yet.

Timonin [2] and Stephanovich [3] suggested the Griffiths scenario. This theory considers poles of the polarization-polarization correlation function and, when applied to PMN, Nb was thought to be the main ferroelectrically active ion. This theory does not take into account the depolarization field, and it considers changes of elastic properties due to the appearance of local polarization. The appearance of local polarization was also considered in Ref. [4]. This study does not exclude that, at $T_d$, a ferroelectric phase transition to the tetragonal phase takes place. This phase transition can be suppressed by random fields. Precursor phenomena have been considered within first-principles computations [5] but they usually take place only in close vicinity of the phase transition and the wide temperature region for such precursors in PMN needs understanding. We will show below that, in order to be consistent with experiment, one has to assume that some kind of a nonpolar isostructural phase transition happens in PMN at $T_d$ and this is critical for the understanding what happens at lower temperatures.

The Burns temperature $T_d$ (in PMN, at about 620 K) [6] marks the appearance of significant deviations of birefringence from a quasilinear temperature dependence. Burns and Dacol connected this deviation with the onset of the formation of PNR's. Neutron scattering measurements have revealed that, below $T_d$, neutron spectra, near the $\Gamma$ point, are dominated by a narrow central peak [7,8]; eigenvectors of the frozen (or slow) displacements [7] are significantly different from the soft mode eigenvectors. Hirota et al [9] proposed to decompose these eigenvectors into the sum of an optical-like (center-of-masses conserving) component and a so-called phase-shift component, describing a uniform shift of all ions in the primitive unit cell.

In this study, we suggest that the nonpolar phase transition at $T_d$ reflects the crossover between the soft-mode and order-disorder dynamics. We assume that the Pb ions have equilibrium positions at the sphere, with the radius $r_0$, surrounding the centrosymmetric Pb position; $r_0$ becomes finite below $T_d$ [10]. Thus, we assume that the Pb ions become off-center, at $T=T_d$. If these off-center ions were not correlated than the phase transition would not happen but rather a smooth transformation would take place. We assume that the Pb off-center ions are correlated over elastic interactions resulting in a cooperative isostructural phase transition. The order parameter of this phase transition could be the average *magnitude* of the Pb displacement. The directions of these displacements are correlated at some finite distances but, we guess, that this correlation is not the driving force of this phase transition. Rather, the elastic interaction is the driving force. However, at lower temperatures, the correlation length of the polarization-polarization correlations increases and results in the formation of PNR's, first, and, then, these PNR's form a spherical glass state [11].

Below, we will formally call the displacements connected with the Pb instability as the Pb displacements although, in fact, the Pb displacements trigger the displacements, polarization and charge redistribution in the group of surrounding ions in the lattice. The Pb instability, in this approach, is caused by local effects, such as the covalent interaction between the Pb and nearest oxygen ions. So, we assume that rather the on-center Pb ionic position becomes unstable at $T_d$ but not the soft optical mode.

## 2. Burns temperature

Due to some randomness in the distribution of the Mg and Nb ions in the lattice of PMN, the Pb ions feel random fields and these fields shift these ions from their centrosymmetric positions [12]. These shifts reach 0.1-0.2 A and are frozen. This is why the distribution function of the Pb displacements has a wide width at any temperature.

It has been shown [7,10] that the Pb displacements do not have a preferable direction in space (spherical model) and, hence, one can introduce a scalar order parameter, $r_0$, corresponding to the magnitude of the Pb displacement. We assume that each Pb ion has two sets of possible states in each primitive unit cell. The first set is on-center or close to the center, and the other set is off-center. The off-center states fill the sphere with the radius $\delta$, with some density. Between the states on the sphere there can be comparatively small barriers. The elastic properties in these two states (on-center and off-center) are assumed different. Hence, we consider two variables, the displacement magnitude, $r$, in each of the primitive cells, and (local) strain (the relative change of the volume) $u$.

The corresponding chemical potentials for the ensembles of independent Pb ions, occupying these two states can be represented in the following form:

$$\begin{aligned}\mu_1(P,T) &= -TS_1(T) + \tfrac{1}{2}a_1(T)r_1^2 + \tfrac{1}{2}C_1 u_1^2 + PV_1(T) \\ \mu_2(P,T) &= -TS_2(T) + \tfrac{1}{2}a_2(T)(r_2 - \delta)^2 + \tfrac{1}{2}C_2 u_2^2 + PV_2(T)\end{aligned} \quad (1)$$

The dependence of the chemical potentials on temperature can be found from anharmonic contributions to Hamiltonian in a mean field approximation by considering a two-well spherical potential fitted by parabolas. We assume an approximately quasilinear temperature dependence of $V_1(T)$ and $V_2(T)$ that corresponds to experiment [13], which we discuss below.

One can exclude the strains and displacements from these expressions by using their equilibrium values. The remaining temperature dependence of these two chemical potentials can be described by monotonous functions. Let these two functions intersect at $T=T_d$ (below we consider the case of zero pressure, for the sake of simplicity):

$$\mu_1(T_d) = \mu_2(T_d) \tag{3}$$

and derivatives of the chemical potentials at this temperature are different due to a large difference in the statistical weight and elastic properties. Close to $T=T_d$, one can represent the difference between the chemical potentials as quasilinear:

$$\mu_1 - \mu_2 = s(T_d - T) \tag{4}$$

We define now the order parameter as the average *magnitude* of the Pb displacements, $r_0$. This magnitude can be found as

$$r_0 = n_1 r_1 + n_2 r_2 \tag{5}$$

where $n_1$ and $n_2$ are the fractions of the two states considered above. Below, we will discuss the conditions for the phase transition of interest. Notice that the independent Pb ions actually do not have any phase transition because of the absence of cooperative phenomena. However, coupling between the Pb displacements can result in a phase transition, at which $r_0$ or its derivative takes a jump.

Consider the effective (temperature dependent) Hamiltonian of a two state model [14-17]:

$$H = s(T - T_d) \sum_i n_i - J \sum_{ij}{}' [2 n_i n_j - n_i - n_j] \tag{7}$$

The meaning of $J$ can be the energy of elastic interactions. Indeed, the change of the local volume of a unit cell results in the appearance of displacements decreasing with the distance as $1/r^2$. This elastic interaction can be sufficient in order to trigger the phase transition of interest.

We considered this Hamiltonian within a mean field approximation and obtained that, at $J > 2k_B T/z$, where $z$ is the number of nearest neighbors, the phase transition is first order [17]. Otherwise, the phase transition does not happen at all and $r_0$ changes with $T$ smoothly, with the main gain in some finite temperature interval (this case can be called as a diffuse first order phase transition). Thus, the Pb instability can result in a (weak) first order phase transition close to second order if the interaction magnitude exceeds the critical one. A pure second order phase transition is also possible if the entropy change at the phase transition is negligible.

The phase transition can be also described by the following Landau expansion with respect to $r_0$:

$$F = \tfrac{1}{2}a(T)r_0^2 + \tfrac{1}{3}br_0^3 + \tfrac{1}{4}cr_0^4 + \tfrac{1}{2}c_L u^2 - \kappa u r_0 \qquad (8)$$

The equilibrium condition is:

$$\tilde{a}(T)r_0 + br_0^2 + cr_0^3 = 0 \qquad (9)$$

where $\tilde{a}(T) = a(T) - \kappa^2/c_L$. The weak first order phase transition (we assume that $b$ is negative and its magnitude is small) happens in the interval $a(T) = [0, b^2/4c]$. Such a phase transition resembles one in the weak crystallization theory [18] and in the theory of coupled rotators [19]. In agreement with this assumption, experiment [20] showed anomalies in the temperature dependence of the heat capacity of PMN in the temperature interval 550-700 K.

Fig. 1 shows the distribution function of the Pb displacements in PMN at different temperatures. This function was experimentally obtained on the basis of treating the intensity of neutron scattering [10]. One can see that, above the Burns temperature, 620 K, the distribution function is one-head whereas, below this temperature, two heads appear, and the distribution function acquires a deep well on-center. Fig. 2 presents the positions of the maxima of the distribution function, $d$, versus temperature. One can see that the behavior of this quantity looks critical at the Burns temperature. The extrapolation of the points near the phase transition provides the exponent 0.3 (the volume of the sphere has 3 times larger exponent). However, the same data can be interpreted as a small jump at the Burns temperature, and, at present, it is difficult to decide which interpretation is right.

The connection of $r_0$ with $u$ should result in an abrupt change of the lattice constant or its temperature derivative, at $T=T_d$. Indeed, experimental data (Fig. 3) show that the lattice parameter of PMN experiences an abrupt deviation from its linear temperature dependence inherent to high temperatures, at $T=T_d$. The same is true for PMN-PT at small concentrations of the PT component. This fact also justifies the inclusion of the elastic energy into the expressions of the chemical potentials.

Besides the interaction of the Pb displacements over local isotropic strains, there is also correlation mediated by phonons. It is obvious that the Pb displacements interact mostly with those phonons, which have the largest Pb component. First principles computations [12] have shown that the Pb displacements are large in the TO1 mode, which corresponds to the displacements of Pb against Nb and O mixed with the displacement of Nb agains O. Other modes containing large (practically 100%) Pb component are TA modes, which are close to the boundary of the Brillouin zone. The latter present practically pure antiphase Pb displacements [12]. As a result of the Pb-off-centering-phonon interaction, the Pb displacements become modulated by the phonons. TO phonons produce polar regions.

The modulation of the Pb displacements can be also considered as a result of coupling between mechanical rotators and phonon modes. It has been shown that this coupling results in a narrow central peak [19] and elastic constants softening with the decrease of temperature [21]. As the dipole's magnitudes and correlation increase towards the freezing temperature, the largest effect is expected at the glass phase transition. Our data also show the abrupt change of the temperature dependence of the lattice constant and birefringence at about 350 K.

Fig. 4 presents the dependence of the refraction index in PMN on temperature [22]. One can see that, at $T=T_d$, the refraction index deviates from its high-temperature behavior and this deviation becomes larger at lower temperatures. The temperature dependence of polarization, which can be obtained from this experiment, just below $T_d$ can be approximated as proportional to $\tau = (T - T_d)/T_d$. Thus, electrostriction provides only quadratic with $\tau$ additions to strain. In contrast to this, the temperature dependence of lattice constant shows an abrupt change of behavior below $T_d$, for PMN. However, below 350 K, one can find correspondence between the change of polarization obtained from birefringence and change of lattice constant obtained by X-Ray diffraction. These data show that, in the temperature interval between 350 K and 620 K, the elastic properties are controlled not only by the change of local polarization but also by other sources, and we connect this difference with the crossover between the soft-mode and order-disorder dynamics. It is obvious that, below $T_d$, the Pb displacements will contribute to hopping dynamics and this hopping should contribute to the phonon peak width, TO-TA interactions, dielectric permittivity dispersion etc.

In our opinion, the eigen vector of the central peak reflects the local distortion appearing due to the Pb displacements to the sphere. This distortion does include, partly, the soft-mode displacements but the main motive behind this distortion is the Pb instability. In order to confirm this statement, we refer to first-principles computations of $KTaO_3$:Li (KLT) [23]. The Li ions, in this solid solution, are known to be shifted from their centrosymmetric positions by about 1 A. In the equilibrium state, these ions have a shift of the center of gravity with respect to the centrosymmetric K positions. The Li barrier is about 1000 K at low temperatures but this barrier rapidly decreases at higher temperatures, and our records do show a significant change of the inclination of inverse dielectric permittivity of KLT at small Li concentration well above the dielectric permittvity peak position that can be considered as an analog of the Burns temperature [24].

First-principles computations [12] revealed a large net Pb displacement in PMN with respect to the centrosymetric Mg positions. The ground state, in this computation, is also complicated by the soft-optical-mode instability, antiferrodistortive instability and large fields due to the difference between the Mg and Nb charges.

We do not consider in the present study self-organizing phenomena in the ensemble of the Pb displacements. These phenomena have been described earlier within the Random bond – Random field model [11].

## 4. Conclusions

The Hirota's et al suggestion [9] about the deconvolution of the frozen (slow) displacements was rather helpful and promising. We have substituted the phase shift introduced by Hirota et al by the shifts triggered by the Pb instability in the oxygen cage. Actually, this shift does lead to the phase shift. The considered model bases on the neutron scattering experiment [10] revealing the change of the Pb displacement distribution function at $T=T_d$ as well as experiment [7] showing isotropy of these displacements. The third significant experiment presents the special eigen vector of the central peak appearing at the Burns temperature [7, 9]. Finally, the abrupt change of the lattice constant behavior at the Burns temperature, which does not follow the change of birefringence, is also an important feature. These principal experiments shed light onto the significance of the Burns temperature in relaxors as the crossover between the soft mode and order-disorder dynamics.

We are grateful for grants 04-02-16103, 05-03-32214 (RFBR) and 01-0105 (INTAS).

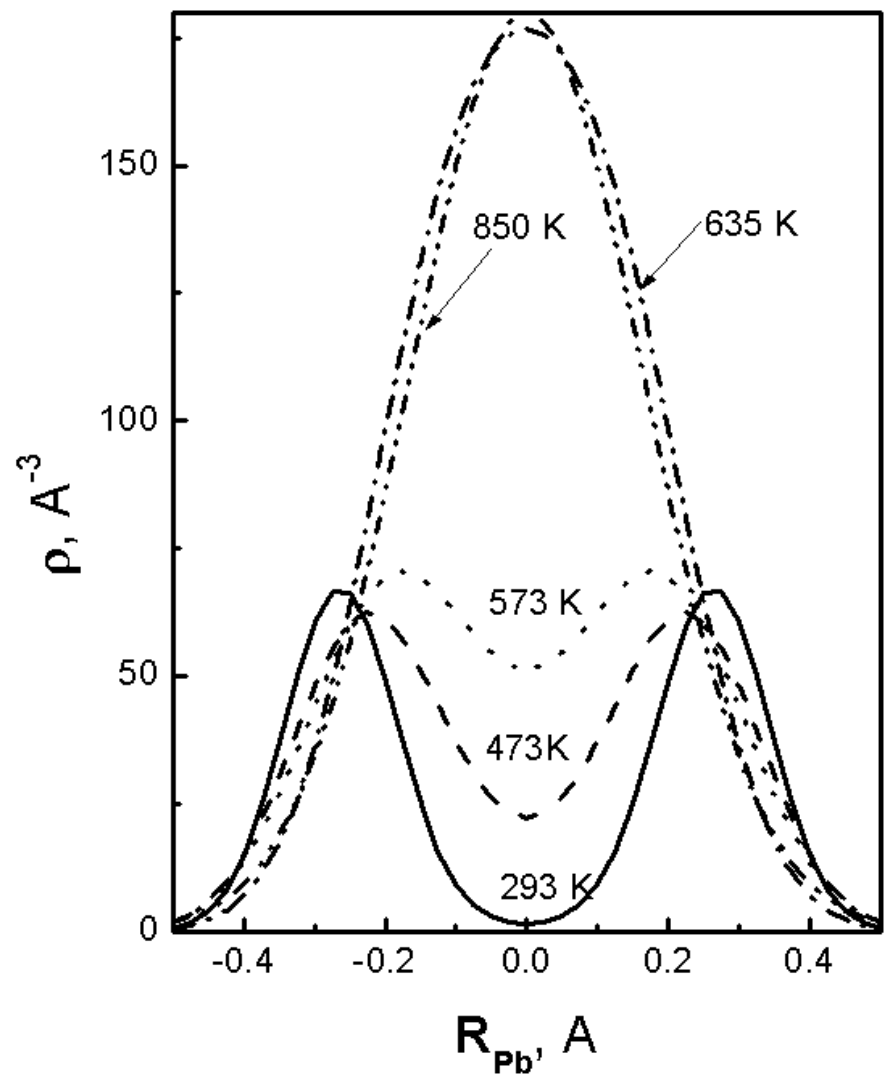

Fig. 1. The distribution of Pb displacements in PMN as it has been extracted from neutron scattering experiments [10].

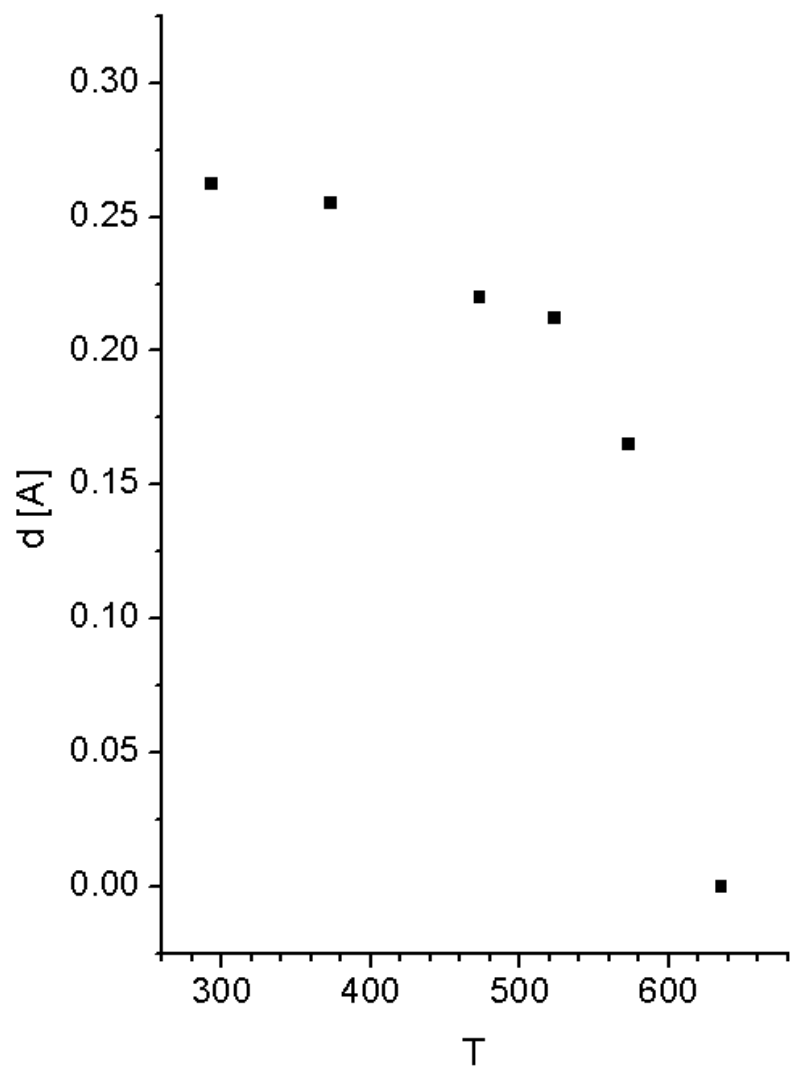

Fig. 2. The dependence of the position of the Pb displacement distribution maximum versus tempearature in PMN.

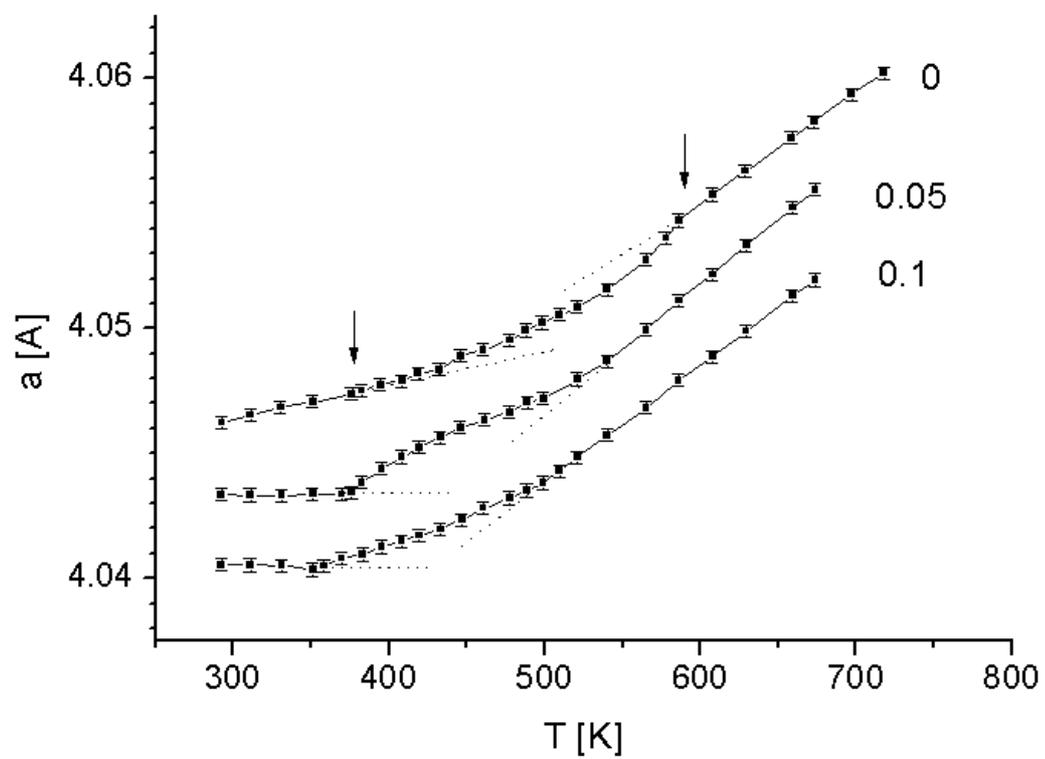

Fig. 3. The dependence of the lattice parameter on temperature in PMN-PT [13]. The numbers show the PT concentration.

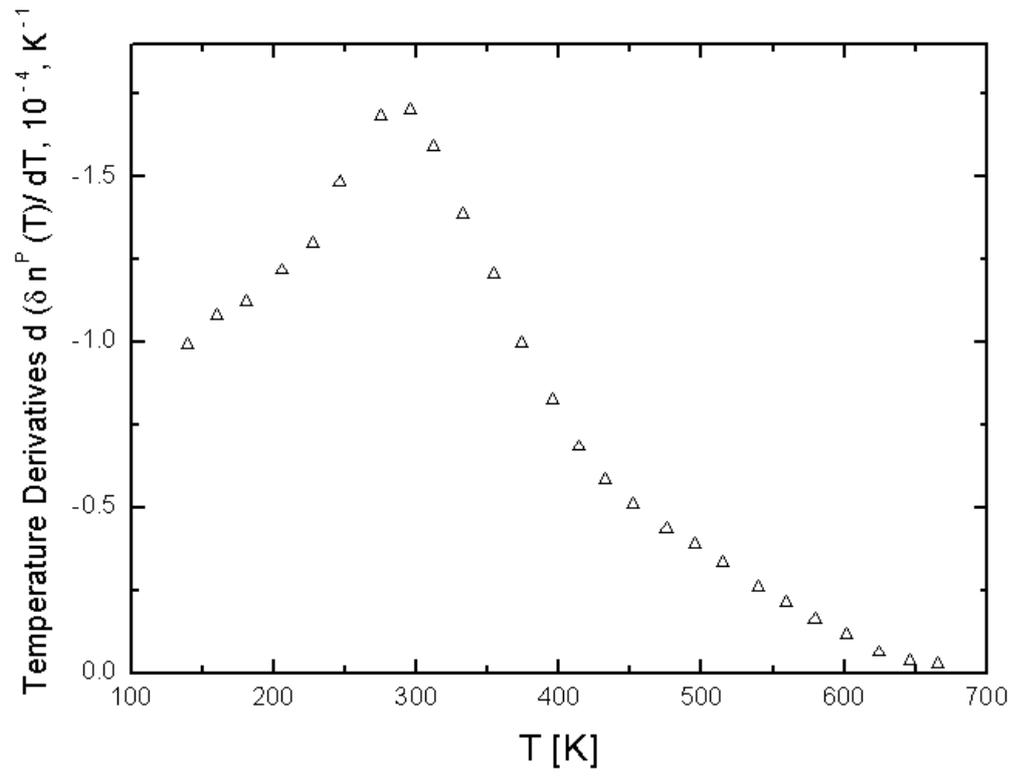

Fig. 4. Temperature derivatives of birefringence for PMN [22].